\documentclass[lettersize,journal]{IEEEtran}
\usepackage[utf8]{inputenc}
\usepackage[T1]{fontenc}
\usepackage{amsmath,amsfonts,amssymb}
\usepackage{algorithmic}
\usepackage{algorithm}
\usepackage{array}
\usepackage{textcomp}
\usepackage{stfloats}
\usepackage{url}
\usepackage{verbatim}
\usepackage{graphicx}
\usepackage{cite}
\usepackage{stfloats}
\usepackage{fancyhdr,float,hyperref,caption, tipa, multirow, subfigure}
\usepackage{mathtools}
\usepackage{cite}
\usepackage{booktabs}
\begin{document}

\title{Single Microphone Own Voice Detection based on Simulated Transfer Functions for Hearing Aids}
\author{
Mathuranathan Mayuravaani, 
W. Bastiaan Kleijn, 
Andrew Lensen, 
Charlotte Sørensen 


}




\maketitle
\begin{abstract}
This paper presents a simulation-based approach to own voice detection (OVD) in hearing aids using a single microphone. While OVD can significantly improve user comfort and speech intelligibility, enabling reliable OVD with a single microphone is desirable for simplifying hardware and reducing power consumption in compact hearing devices. However, most existing solutions rely on multiple microphones or additional sensors, increasing device complexity and cost. To enable ML-based OVD without requiring costly transfer-function measurements, we propose a data augmentation strategy based on simulated acoustic transfer functions (ATFs) that expose the model to a wide range of spatial propagation conditions. A transformer-based classifier is trained using analytically generated ATFs and further fine-tuned using numerically simulated ATFs with increasing geometric realism. This hierarchical adaptation enables the model to refine its spatial understanding while maintaining generalization. Experimental results show 95.52\% accuracy on simulated head-and-torso test data and 90.02\% accuracy for one-second speech segments, demonstrating robustness to short durations. When evaluated on real-world hearing-aid recordings, few-shot fine-tuning using a small subset achieves 91\% accuracy, demonstrating that limited real-world data can effectively adapt the simulation-trained model. These results highlight the potential of simulation-based training for enabling practical single-microphone OVD systems in hearing aids.
\end{abstract}

\section{Introduction}

\IEEEPARstart{H}{earing aids} play a crucial role in improving the quality of life for individuals with hearing impairments by amplifying speech. However, a common complaint among hearing aid users is that their own voice sounds excessively loud or unnatural when using the device~\cite{froehlich2018perception}. To mitigate this issue, hearing care professionals often reduce amplification levels to enhance comfort. This reduction, however, compromises speech clarity and audibility, making communication difficult~\cite{hoydal2017new}. Developing a robust method for efficiently identifying the wearer’s own voice is crucial, as it enables personalized adjustments to maintain comfort and intelligibility~\cite{hengen2020perception}.

Although many modern hearing aids incorporate multiple microphones, reliable single microphone own voice detection (OVD) remains desirable for several practical reasons. Multi-microphone processing increases hardware cost, power consumption, and calibration complexity, limiting applicability in low-cost devices \cite{kates2008hearing}. In addition, some users rely on a single device due to unilateral hearing loss or personal preference. Moreover, signals may become unavailable or degraded due to occlusion, wind noise, or power-saving modes \cite{chung2004challenges}, making single-microphone OVD a useful fallback. For these reasons, single-microphone OVD remains a practically important and complementary alternative to multi-microphone solutions.

The majority of existing OVD methods rely on traditional signal processing techniques, such as cross-correlation, adaptive filtering, and beamforming~\cite{merks2014hearing, al2014cross, rasmussen2009method, zohourian2017binaural}. These methods typically require multiple microphones and complex directional filtering, which increases hardware costs and makes them unsuitable for individuals with single-ear hearing loss.

Machine learning (ML) approaches, often applied in single-microphone configurations, have gained popularity in speaker identification and distance estimation tasks~\cite{wang2023lightweight, patterson2022, wang2023target, neri2024speaker}, as they can capture complex patterns in speech signals and improve adaptability. However, many ML-based approaches rely primarily on speaker-dependent vocal characteristics to distinguish the user’s voice. In contrast, this work proposes a spatial-analysis-based approach that uses differences in acoustic propagation paths between the user’s own voice and external speakers.
Acoustic transfer functions (ATFs) are used to model the distinct propagation paths of own voice and external speech as they reach the hearing-aid microphone, and a data-driven classifier is trained to learn discriminative spatial–spectral patterns from ATF-augmented speech.

To overcome the practical difficulty of collecting large-scale measured ATFs across anatomies and device configurations, we propose a simulation-based ATF generation pipeline that progressively increases geometric realism. We first generate analytically derived ATFs based on simplified geometries to cover a wide range of spatial configurations, enabling scalable data generation with controlled variation. We then generate numerically simulated ATFs using boundary-element modeling, transitioning from a rigid sphere to a detailed head-and-torso model. These ATFs are used to augment speech signals for training the OVD classifier. We formulate OVD as a segment-level binary classification and use a transformer-based classifier to aggregate frame-level features into a single decision.

The main contributions of this work are as follows: (i) enabling scalable training of single-microphone OVD models using simulated ATFs without requiring costly measured transfer functions; (ii) employing a progressive training strategy that transitions from analytical to numerically simulated ATFs with increasing anatomical realism; and (iii) demonstrating that spatial cues embedded through simulated ATFs enable effective single-microphone OVD on real-world hearing-aid recordings. This study focuses on offline segment-level detection as a feasibility analysis; causal streaming implementation with low-latency inference is left for future work.

\vspace{-1mm}
\section{Background}
OVD has been a longstanding challenge in speech processing, particularly for hearing aids. Early approaches relied on classical signal processing techniques, later evolving to incorporate multichannel methods and, more recently, ML-based solutions. One of the earliest OVD methods was proposed by Rasmussen et al.~\cite{rasmussen2009method}, who used a cross-correlation technique to exploit the symmetrical positioning of the user's mouth. Their method aimed to detect speech originating from the median plane. To improve near-field detection, they introduced the mouth-to-random-far-field (M2R) technique; however, if a straight-ahead external speaker is close enough, their voice may also exhibit near-field characteristics, posing a challenge for accurate differentiation. Another classical approach involved adaptive filtering, as proposed by Merks et al.~\cite{merks2014hearing}, where two microphones were used to estimate a transfer function for OVD. While effective in some cases, this method was susceptible to errors in the presence of strong external speech sources. Bone-conductive sensors have also been explored~\cite{merks2014hearing, wei2022self}, but their effectiveness is limited by low signal-to-noise ratio (SNR) and the need for direct skull contact.

 Hoang et al.~\cite{hoang2022multichannel} applied beamforming and relative ATFs for interfering speech suppression. They considered the ATF from the user’s mouth to the microphones to be approximately time-invariant. With the microphones positioned close to the user’s mouth, the own voice signal can be louder than both the target and interfering speech, especially at lower frequencies when the own voice is active. However, these considerations may not apply to a speaker facing directly forward or with varying loudness. Frequency-dependent variations in the response of rear microphones in head-mounted arrays have been observed and attributed to diffraction and directional propagation of own voice~\cite{kazama2024measurement}. A similar trend appears in our free-space ATFs, where the response changes progressively with frequency (see Section~\ref{sec:basic_formulas}).

ML-based OVD methods have also been explored. Pohlhausen et al.~\cite{pohlhausen2022near} proposed a binaural OVD method utilizing near-ear microphones and a random forest classifier, distinguishing own voice from external sounds based on phase and amplitude symmetry. Similarly, Pertilä et al.~\cite{pertila2021online} developed an LSTM-based online OVD system for in-ear devices using multiple microphones and accelerometers, which increase system complexity and cost. Lopez-Espejo et al.~\cite{lopez2019keyword, lopez2020improved} trained a deep residual network for OVD within a keyword spotting system, leveraging measured head-related transfer functions (HRTFs) and own voice transfer functions (OVTFs) from a hearing assistive device. However, these approaches rely on measured, device-specific transfer functions, which limits scalability because collecting such measurements across different anatomies, device placements, and acoustic conditions is costly and time-consuming. As a result, the coverage of possible acoustic propagation paths remains limited. This motivates the use of simulation-based ATF generation, which enables scalable training data generation with controlled variation across spatial configurations. Additionally, distance-based ML approaches for speaker separation have been investigated using single-channel audio~\cite{patterson2022, neri2024speaker}.

 Acoustic modeling has played a significant role in understanding voice propagation and directivity patterns. Various analytical models~\cite{blandin2019influence, brandner2020pilot} have been employed to study speech and singing sound propagation, often utilizing simplified representations of the human head. Many of these models approximate the acoustic effects of the head and mouth using geometrical shapes, such as spheres. For instance, Blandin et al.~\cite{blandin2019influence} analyzed the directivity patterns around the head, assuming a vibrating plane piston model set within an infinite baffle. Their work aimed to enhance the quality of artistic performances in concert halls by investigating the influence of the mouth opening size on voice directivity, comparing both measurements and simulations. They observed that the overall features of the radiated sound could be predicted from the properties of the vibrating piston. Similarly, the impact of mouth configuration and torso shape on voice directivity was explored during singing~\cite{brandner2020pilot}. Utilizing both piston and spherical cap models for their simulations, they found that significant changes in the mouth opening size considerably affect voice directivity. In our approach, we present a model that represents the human head as a rigid sphere with the mouth modeled as a vibrating cap on its surface, providing a physically grounded approach to detecting the wearer's own voice in a hearing aid. A point source is represented as an external speaker, allowing us to analyze how sound scatters from the rigid sphere. By jointly modeling both the wearer’s voice and external sources, we capture the spatial propagation differences that distinguish them, allowing us to generate ATFs that enable us to augment the speech data to train the ML algorithm.

While previous works have applied ML for OVD and utilized transfer functions in speech modeling, to the best of our knowledge, no study has incorporated analytical and numerical ATFs for augmenting training data in ML-based OVD using a single microphone.
In this work, we propose a data augmentation strategy that enhances training data diversity using synthetic ATF-augmented speech, and we evaluate a conformer encoder-based classifier~\cite{gulati2020conformer} for segment-level own-voice versus external-speaker classification. The conformer architecture provides a flexible mechanism for aggregating frame-level speech features into a single decision, improving robustness to phonetic variability while operating in a single microphone setting.

\vspace{-2mm}

\section{Method}

In this study, our objective is to classify the hearing aid wearer’s own voice versus external speakers' voices. To achieve this, we employ an ML algorithm to perform the classification. Since ML models require large and diverse training datasets to generalize effectively, we augment the speech data using ATFs. Specifically, we generate training samples using both analytically derived ATFs, which introduce controlled variability, and numerically simulated ATFs, which incorporate detailed human anatomy for more realistic spatial representations. The model was first trained on analytically augmented data and subsequently fine-tuned using samples augmented from numerical simulations.

To construct these ATFs for data augmentation and analysis, we model the wearer’s head as a rigid sphere that captures scattering effects shaping the acoustic paths to the hearing-aid microphone. External speech is modeled as radiation from a point source scattered by the rigid sphere, whereas own voice is modeled as radiation from a vibrating spherical cap on the same sphere, representing mouth radiation coupled with head scattering. In both cases, the resulting pressure is evaluated at the hearing-aid microphone location on the sphere surface. The same source definitions are used in the numerical simulations, while the geometry is progressively refined from a rigid sphere to detailed head and head-and-torso models to increase anatomical realism. Similar geometric approximations, such as spherical caps and pistons, have been used in previous studies~\cite{blandin2019influence, brandner2020pilot, aarts2011comparing} to model sound radiation and head-related scattering.  

Section~\ref{sec:basic_formulas} presents the fundamental equations governing the considered geometries. Section~\ref{sec:ATF_modeling} details the analytical approach for modeling sound propagation. We validate this approach by comparing analytical results with numerical simulations and also generate ATFs using simulation software, as discussed in Section~\ref{sec:numerical_modeling}. Finally, the OVD algorithm is described in Section~\ref{sec:ovd_algorithm}.
\vspace*{-4mm}
\subsection{Geometry and basic formulas}
\label{sec:basic_formulas}
\begin{figure}[H]
    \centering
    \includegraphics[trim=342 212 365 210, clip, scale=0.64]{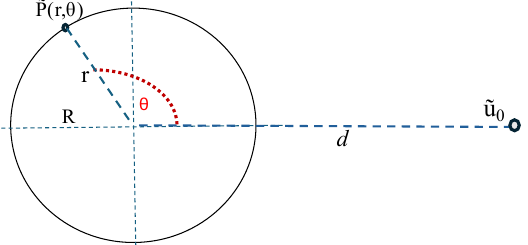}
    \caption{Geometry of point source and rigid sphere.}
    \label{fig:fig16}
\end{figure}

We consider a point source with a volume velocity $\tilde{U}_{0}$ located at a distance $d$ from a rigid sphere of radius $R$ (Fig.~\ref{fig:fig16}) to model the effect of an external speaker's voice reaching a hearing-aid user. Our goal is to determine the pressure field $\tilde{p}(\mathbf{r}, \theta)$ at an observation point on the sphere, corresponding to the ear location of the hearing-aid user. $r_1 = \sqrt{r^2 + d^2 - 2rd \cos \theta}$ is the distance from the point source to the observation point. In spherical coordinates, the incident pressure field due to the point source is~\cite{beranek2019acoustics, ansolpoint}:
\begin{equation}
\label{eq:pointsource}
    \tilde{p}(r, \theta) = jk\rho_{0}c\tilde{U_0}\frac{e^{-jkr_1}}{4\pi r_1},  
\end{equation}
where $j$ is the imaginary unit, $k$ is the wave number, $\rho_{0}$ is the density of air and $c$ is the speed of sound in air. 

\vspace{-1mm}
This expression describes the free-field pressure at an arbitrary location before considering reflections and scattering. This expression can be expanded in terms of spherical Bessel functions $j_n$ and Legendre functions $P_n$ as follows~\cite{beranek2019acoustics}:

\begin{multline}
\label{eq:incident_expansion}
    \tilde{p}_{I}(r,\theta) = \frac{k^2 \rho_0 c \tilde{U}_0}{4 \pi} \sum_{n=0}^{\infty} (2n + 1) h_{n}^{(2)}(kd)j_n(kr)P_{n}(\cos\theta) , \\
    r \le d.
\end{multline}

When an external speaker’s voice reaches the hearing-aid user, part of the sound is scattered by the head. The scattered pressure field is given by~\cite{beranek2019acoustics}:
\par\vspace*{-3mm}

\begin{multline}
\label{q:pss}
    \tilde{p}_s(r, \theta) =  -\frac{k^2 \rho_0 c \tilde{U}_0}{4 \pi}\\ \sum_{n=0}^{\infty} (2n+1) h_{n}^{(2)}(kd) \frac{j^{'}_n(kR)}{ h^{'(2)}_n(kR)} h^{(2)}_n(kr) P_{n}(\cos\theta).
\end{multline}

\noindent The total pressure field at the observation point is obtained by summing the incident and scattered fields:
\begin{equation}
\label{eq:total_pressure}
	\tilde{p}(r,\theta) = \tilde{p}_I(r,\theta) + \tilde{p}_{s}(r,\theta).
\end{equation}

\begin{figure}[H]
    \centering
    \includegraphics[trim=359 220 465 200, clip, scale=0.64]{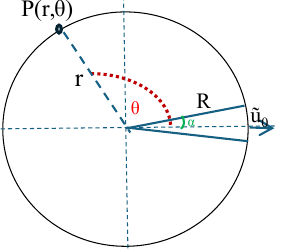}
    \caption{Geometry of oscillating cap in a rigid sphere.}
    \label{fig:fig17}
\end{figure}

This total pressure field is critical for analyzing how an external speaker’s voice propagates to the hearing-aid user. In contrast to an external speaker, the hearing-aid user's own voice originates from the mouth and propagates differently. To model this, we represent the mouth as a vibrating spherical cap on a rigid sphere that vibrates with an axial velocity of $\tilde{u}_{0}$ (Fig.~\ref{fig:fig17})~\cite{beranek2019acoustics, aarts2010sound}. The near-field pressure due to the oscillating cap describes the propagation of the user’s own voice from the mouth to the hearing-aid microphone. This pressure field is given by~\cite{beranek2012acoustics}:

\begin{equation}
\label{q:equation13}
\resizebox{\columnwidth}{!}{$ 
\begin{aligned}
	\tilde{p}(r,\theta) &= -jk \rho_0 c\tilde{u}_0 \bigg( 
    \frac{\sin^2\alpha}{4h_{0}^{'(2)}(kR)}h_{0}^{(2)}(kr) 
    \quad +  \frac{1 - \cos^3\alpha}{2h_{1}^{'(2)}(kR)} h_{1}^{(2)}(kr)\cos\theta \\
	&\quad -\sin\alpha \sum_{n=2}^{\infty} (2n+1) 
    \frac{\sin\alpha P_n(\cos\alpha) + \cos\alpha P^{1}_n(\cos\alpha)}
    {2(n-1)(n+2) h_{n}^{'(2)}(kR)} h_{n}^{(2)}(kr)P_n(\cos\theta) \bigg),
\end{aligned}
$}
\end{equation}

\noindent where $\alpha$ is the half-angle of the arc formed by the cap.

\begin{figure}[!b]
\begin{center}
	\centering
	 \includegraphics[trim=0.8cm 0.55cm 0.5cm 0.55cm, clip, scale=0.28]{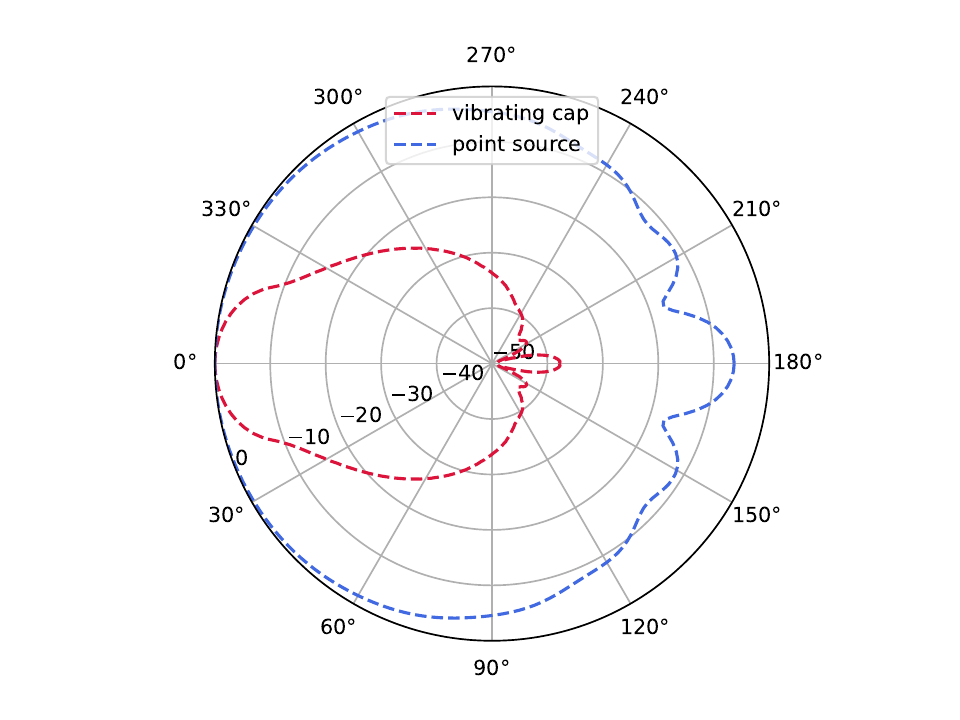}
     \caption{Directivity pattern $20 \log_{10}( D(\theta)/D(0))$  of the near-field pressure for a straight-ahead point source scattering from a 7 cm rigid sphere and a spherical cap ($\alpha=20^\circ$).}

     \label{fig:fig18}
 \end{center}
\end{figure}

\begin{figure*}[!h]
\begin{center}
	\centering
    \includegraphics[trim=200 230 235 220, clip, scale=0.8]{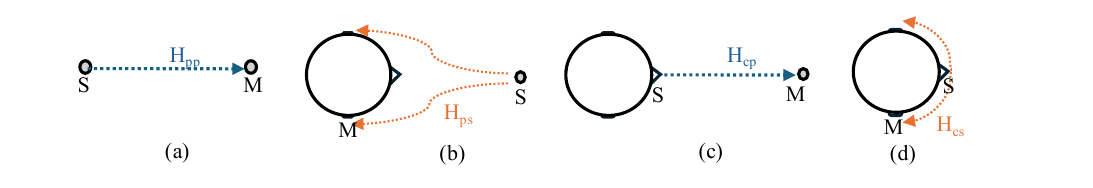}
	\caption{Mapping the pressure between (a) a point source and a point receiver, (b) a point source and a rigid sphere, (c) a rigid sphere source with vibrating cap and a point receiver, and (d) vibrating spherical cap and location on the rigid sphere. Here, S denotes the source location and M denotes the receiver (microphone) location.}
    \label{fig:fig2}
 \end{center}
\end{figure*}
\begin{figure*}[b!]
\begin{center}
	\centering
	\includegraphics[trim=150 255 300 200, clip, scale=0.85]{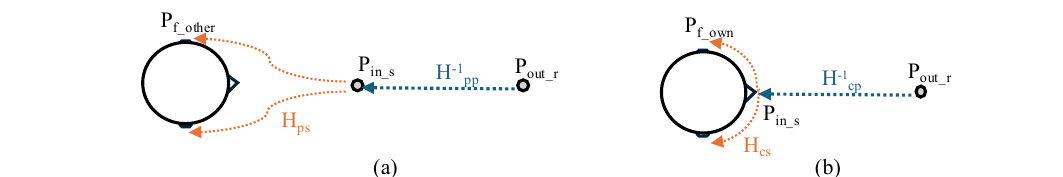}
	\caption{Mapping the pressure at an angle on the sphere  when (a) a speech signal is from a distant point source (external speaker) and (b) a speech signal is from a vibrating cap on the rigid sphere (own voice).}
\label{fig:fig3}
 \end{center}
\end{figure*}
\vspace{-3mm}

A comparison of directivity patterns at 5~kHz is presented in Fig.~\ref{fig:fig18}, considering a point source scattering from a rigid sphere and a vibrating spherical cap on the sphere. The results show a significant difference in the directivity pattern with respect to the on-axis direction. This effect becomes more pronounced at higher frequencies, leading to reduced pressure at the ear location in the own-voice scenario.

\begin{figure*}[h!]
    \centering

    \begin{minipage}{.5\textwidth}
        \centering
        \subfigure[Analytical results.]{
            \includegraphics[scale=0.47]{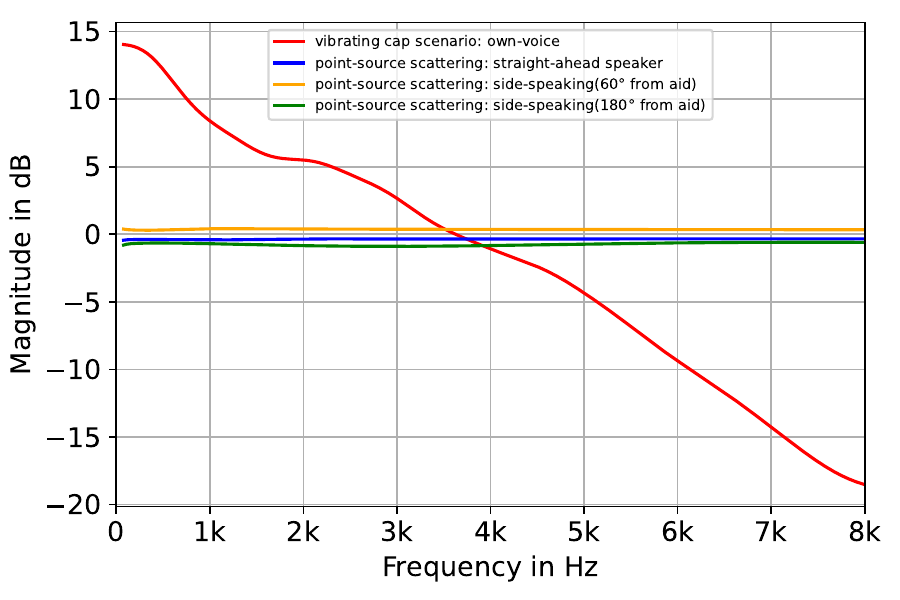}}
    \end{minipage}%
    \begin{minipage}{.5\textwidth}
        \centering
        \subfigure[Simulation results.]{
            \includegraphics[trim=0.2cm 0.25cm 29.5cm 0.25cm, clip, scale=0.64]{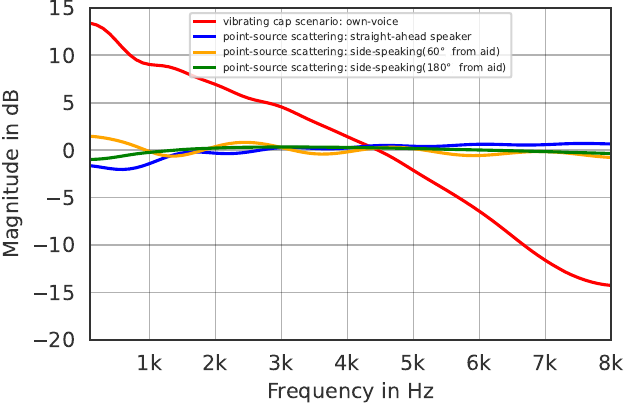}}
    \end{minipage}%
    \vspace{-1mm}
    \caption{Comparison between analytical and numerical simulation results of ATF with rigid sphere.}
    \label{fig:fig24}
\end{figure*}  

\subsection{Analytical Modeling Approach}
\label{sec:ATF_modeling}

This section describes how analytically derived ATFs are computed and applied to generate synthetic hearing-aid microphone signals. We consider two classes: (i) external-speaker and (ii) own-voice. The two classes differ in the assumed geometry model and propagation path: external speech is modeled as radiation from a point source scattered by a rigid sphere, whereas own voice is modeled as radiation from a vibrating spherical cap mounted on the same sphere. The resulting ATFs characterize the frequency-domain mapping from the source model to the hearing-aid microphone position and are subsequently used to generate ATF-augmented speech samples for ML training (described in Section~\ref{sec:ovd_algorithm}).

Let $P_{out\_r}(k)$ denote the complex pressure spectrum at a reference microphone receiver location (used as a known observation in the simulation pipeline). Our objective is to generate the pressure spectrum at the hearing-aid location, denoted by $P_{f\_other}(k)$ and $P_{f\_own}(k)$, for both external and own voice conditions. The mapping is performed in two steps. First, the ATF corresponding to the reference propagation path is inverted to estimate an equivalent source spectrum. Second, the ATF from the source to the hearing-aid microphone is applied to obtain the pressure at the hearing-aid. We define the following frequency-domain transfer functions parameterized by the acoustic wavenumber $k=\omega/c$ (see Fig.~\ref{fig:fig2}):

\begin{itemize}
    \item $H_{pp}(k)$: ATF from a point source to a point receiver. 
    \item $H_{ps}(k)$: ATF from a point source to a receiver location on the rigid sphere.
    \item $H_{cp}(k)$: ATF from a vibrating spherical-cap source (own-voice model) to a point receiver. 
    \item $H_{cs}(k)$: ATF from a vibrating spherical-cap source to a receiver location on the rigid sphere. 
\end{itemize}
For the external-speaker class, the speech waveform is assumed to radiate from a point source. The pressure at the reference receiver $P_{out\_r}(k)$ is first mapped back to an estimate of the source pressure using the inverse ATF $H_{pp}^{-1}(k)$, and then forward-propagated to the hearing-aid microphone position using $H_{ps}(k)$ (see Fig.~\ref{fig:fig3}(a)):

\begin{equation}
P_{f\_other}(k) = H_{ps}(k)\, H_{pp}^{-1}(k)\, P_{out\_r}(k).
\end{equation}

For the own voice class, the speech waveform is assumed to radiate from a vibrating spherical cap mounted on a rigid sphere, representing sound radiation from the wearer's mouth. The reference observation is mapped to an equivalent cap-source spectrum using $H_{cp}^{-1}(k)$ and then propagated to the hearing-aid microphone using $H_{cs}(k)$ (see Fig.~\ref{fig:fig3}(b)):
\begin{equation}
P_{f\_own}(k) = H_{cs}(k)\, H_{cp}^{-1}(k)\,P_{out\_r}(k).
\end{equation}

\noindent These ATFs (see Fig.~\ref{fig:fig24}(a)) enable the simulation of how speech propagates to the hearing-aid user.

\subsubsection{$H_{pp}(k)$}
\label{sec:Hpp}

The ATF models free-field propagation from a point source to a point receiver (Fig.~\ref{fig:fig2}(a)). From the point-source pressure expression in \eqref{eq:pointsource}, the corresponding ATF is defined as:

\begin{equation}
   H_{pp}(k) = \frac{r_0 e^{-jkr_1} }{ r_1 e^{-jkr_0}}.
   \label{q:pointsourceATF}
 \end{equation}
Here, $r_0$ is a reference distance close to the source, and $r_1$ is the source-to-receiver distance. We use $H_{pp}^{-1}(k)$ to map the reference pressure back to the equivalent source location.

\begin{figure*}[!b]
    \centering
    \begin{minipage}{.32\textwidth}
        \centering
        \subfigure[3D Rigid sphere model.]{
            \includegraphics[scale=0.49]{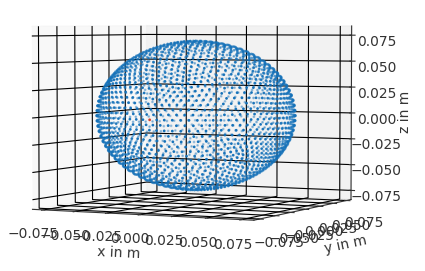}}
    \end{minipage}%
    \begin{minipage}{.32\textwidth}
        \centering
        \subfigure[3D Human head model.]{
            \includegraphics[scale=0.32]{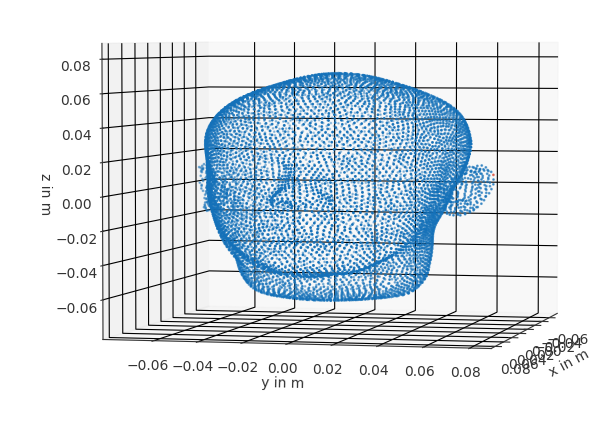}}
    \end{minipage}%
    \begin{minipage}{.32\textwidth}
        \centering
        \subfigure[3D Human head-and-torso model.]{
            \includegraphics[scale=0.44]{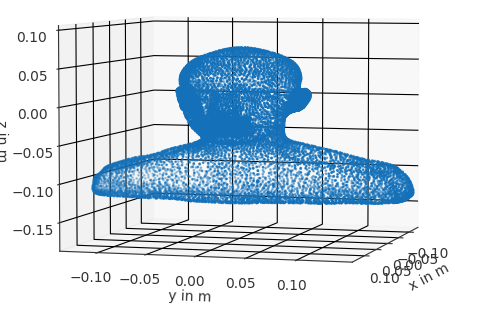}}
    \end{minipage}
       \caption{3D model meshes created in Blender for simulation.}
    \label{fig:fig15}
\end{figure*}

\subsubsection{$H_{ps}(k)$}
\label{sec:Hps}
The ATF models propagation from a point source to a receiver location on a rigid sphere (Fig.~\ref{fig:fig2}(b)). It is defined as the ratio of the total pressure at the receiver on the sphere surface, obtained from \eqref{eq:total_pressure} by setting $r=R$, to the corresponding free-field point-source pressure at distance $r_0$, given in \eqref{eq:pointsource}. This reference pressure corresponds to the incident free-field pressure in the absence of the sphere, which can be expressed by \eqref{eq:pointsource} or equivalently by \eqref{eq:incident_expansion}. With this formulation, $H_{ps}(k)$ captures both propagation and rigid sphere scattering:

\begin{equation}
\resizebox{\columnwidth}{!}{$ %
    H_{ps}(k) =
    \frac{
    k r_0
    \sum_{n=0}^{\infty} (2n + 1) h_{n}^{(2)}(kd)
    \left(
    j_n(kR) -
    \frac{j^{'}_n(kR)}{h^{'(2)}_n(kR)}
    h^{(2)}_n(kR)
    \right)
    P_{n}(\cos\theta)
    }{
    j e^{-jkr_0}
    }.
$}
\label{eq:Hps}
\end{equation}

\subsubsection{$H_{cp}(k)$}
\label{sec:Hcp}
The ATF models propagation from a vibrating spherical-cap source mounted on a rigid sphere to a point receiver (Fig.~\ref{fig:fig2}(c)). It is derived from the spherical-cap pressure field in \eqref{q:equation13} by expressing the pressure at the receiver distance $r_1$ relative to that at the reference distance $r_0$, which is chosen close to the cap. With this formulation, $H_{cp}(k)$ captures the propagation from the vibrating spherical-cap source to the point receiver:

\begin{equation}
\resizebox{\columnwidth}{!}{$ %
   H_{cp}(k) = \frac{\bigg(\splitdfrac{\frac{\sin^2\alpha}{4h_{0}^{'(2)}(kR)}h_{0}^{(2)}(kr_{1}) + \frac{1 - \cos^3\alpha}{2h_{1}^{'(2)}(kR)} h_{1}^{(2)}(kr_1)
	-\sin\alpha }{ \sum_{n=2}^{\infty} (2n+1) \frac{\sin\alpha P_n(\cos\alpha) + \cos\alpha P^{1}_n(\cos\alpha)}{2(n-1)(n+2) h_{n}^{'(2)}(kR)} h_{n}^{(2)}(kr_1)} \bigg)} {\bigg(\splitdfrac{\frac{\sin^2\alpha}{4h_{0}^{'(2)}(kR)}h_{0}^{(2)}(kr_{0}) + \frac{1 - \cos^3\alpha}{2h_{1}^{'(2)}(kR)} h_{1}^{(2)}(kr_0) 
	-\sin\alpha}{ \sum_{n=2}^{\infty} (2n+1) \frac{\sin\alpha P_n(\cos\alpha) + \cos\alpha P^{1}_n(\cos\alpha)}{2(n-1)(n+2) h_{n}^{'(2)}(kR)} h_{n}^{(2)}(kr_0)} \bigg)}.
$}
\end{equation}

\noindent We use $H_{cp}^{-1}(k)$ to estimate the equivalent spherical-cap source spectrum from the reference observation.

\subsubsection{$H_{cs}(k)$}
\label{sec:Hcs}

The ATF models propagation from a vibrating spherical-cap source to a receiver location on the rigid sphere (Fig.~\ref{fig:fig2}(d)), representing how the own voice is observed at the hearing-aid microphone. It is derived from the spherical-cap pressure field in \eqref{q:equation13} by evaluating the pressure on the sphere surface and expressing the response at the receiver angle $\theta$ relative to the corresponding on-axis reference response.  With this formulation, $H_{cs}(k)$ captures the directional own-voice radiation from the vibrating spherical cap to the hearing-aid microphone location on the sphere surface:

\begin{equation}
\resizebox{\columnwidth}{!}{$ %
   H_{cs}(k) = \frac{ \bigg(\splitdfrac{\frac{\sin^2\alpha}{4h_{0}^{'(2)}(kR)}h_{0}^{(2)}(kR) + \frac{1 - \cos^3\alpha}{2h_{1}^{'(2)}(kR)} h_{1}^{(2)}(kR)\cos\theta 
	-\sin\alpha }{ \sum_{n=2}^{\infty} (2n+1) \frac{\sin\alpha P_n(\cos\alpha) + \cos\alpha P^{1}_n(\cos\alpha)}{2(n-1)(n+2) h_{n}^{'(2)}(kR)} h_{n}^{(2)}(kR) P_n(\cos\theta)} \bigg)} {\bigg(\splitdfrac{\frac{\sin^2\alpha}{4h_{0}^{'(2)}(kR)}h_{0}^{(2)}(kR) + \frac{1 - \cos^3\alpha}{2h_{1}^{'(2)}(kR)} h_{1}^{(2)}(kR)
	-\sin\alpha}{ \sum_{n=2}^{\infty} (2n+1) \frac{\sin\alpha P_n(\cos\alpha) + \cos\alpha P^{1}_n(\cos\alpha)}{2(n-1)(n+2) h_{n}^{'(2)}(kR)} h_{n}^{(2)}(kR) } \bigg)}. 
$}
  \end{equation}

\begin{figure*}[hbt!]
    \centering
    \begin{minipage}{.33\textwidth}
        \centering
        \subfigure[Rigid sphere at varying angles.]{
            \includegraphics[trim=0.2cm 0.25cm 33cm 0.25cm, clip, scale=0.36]{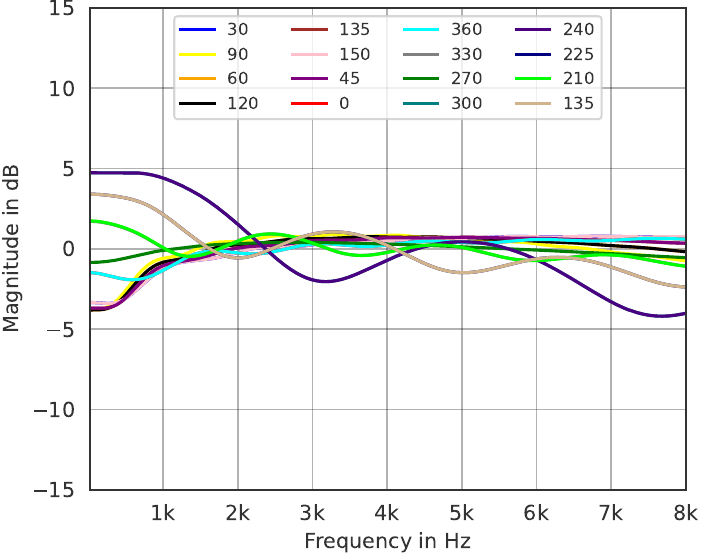}}
    \end{minipage}%
   \begin{minipage}{.33\textwidth}
        \centering
        \subfigure[Head at varying angles.]{
            \includegraphics[trim=0.2cm 0.25cm 33cm 0.25cm, clip, scale=0.36]{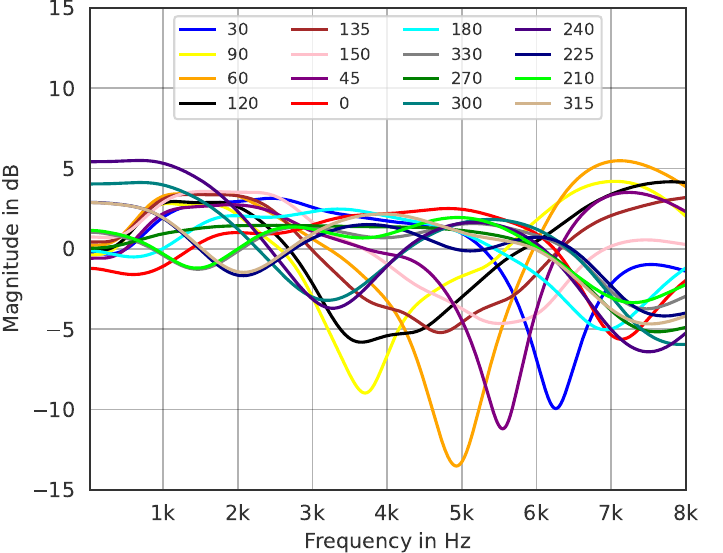}}
    \end{minipage}%
     \begin{minipage}{.33\textwidth}
        \centering
        \subfigure[Head-and-torso at varying angles.]{
           \includegraphics[trim=0.2cm 0.25cm 33cm 0.25cm, clip, scale=0.36]{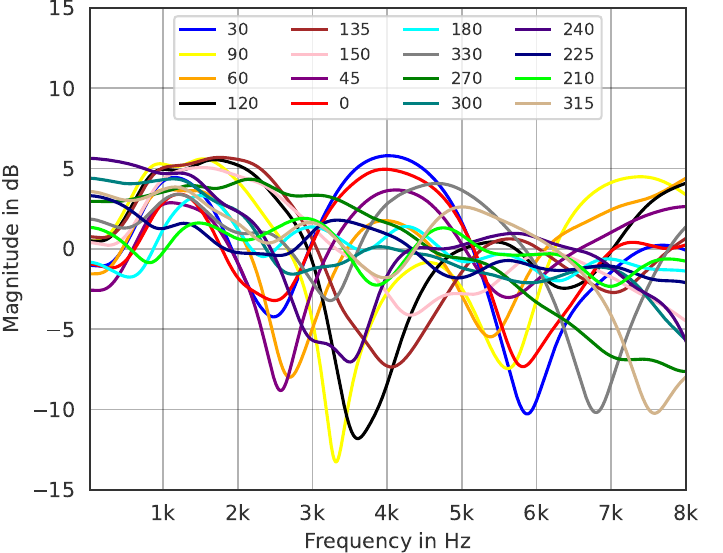}}
    \end{minipage}%

     \begin{minipage}{.33\textwidth}
        \centering
        \subfigure[Rigid sphere with varying radii.]{
            \includegraphics[trim=0.2cm 0.25cm 33cm 0.25cm, clip, scale=0.36]{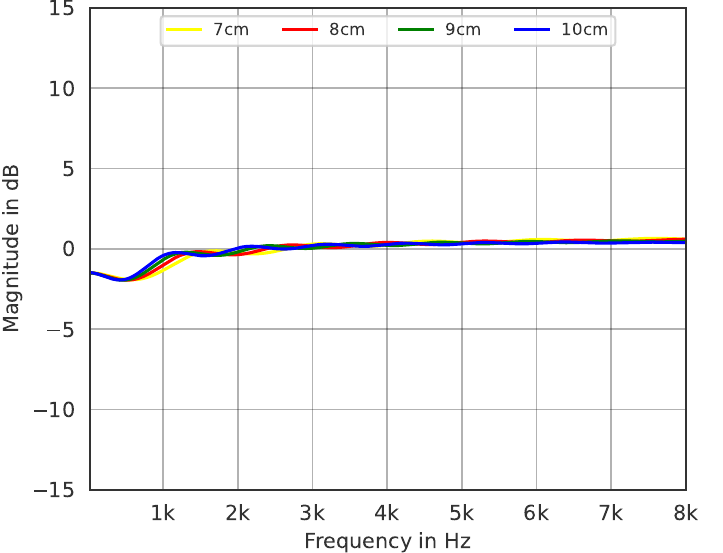}}
    \end{minipage}%
    \begin{minipage}{.33\textwidth}
        \centering
        \subfigure[Head with varying radii.]{
            \includegraphics[trim=0.2cm 0.25cm 33cm 0.25cm, clip, scale=0.36]{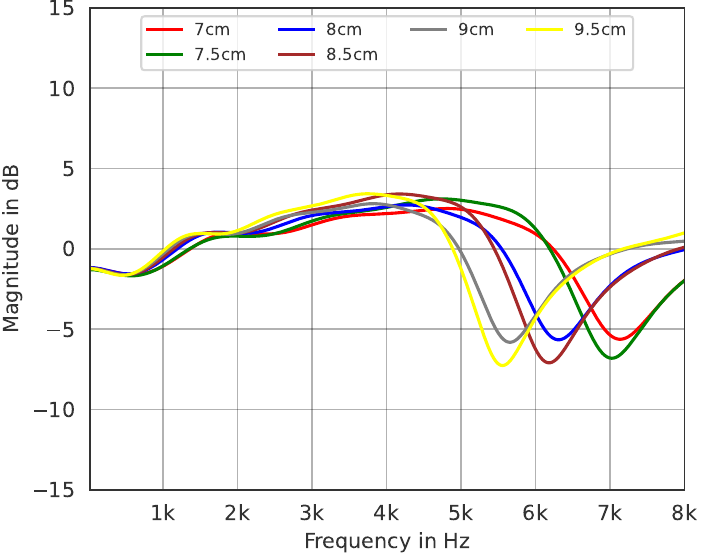}}
    \end{minipage}%
     \begin{minipage}{.33\textwidth}
        \centering
        \subfigure[Head-and-torso with varying radii.]{
            \includegraphics[trim=0.2cm 0.25cm 33cm 0.25cm, clip, scale=0.36]{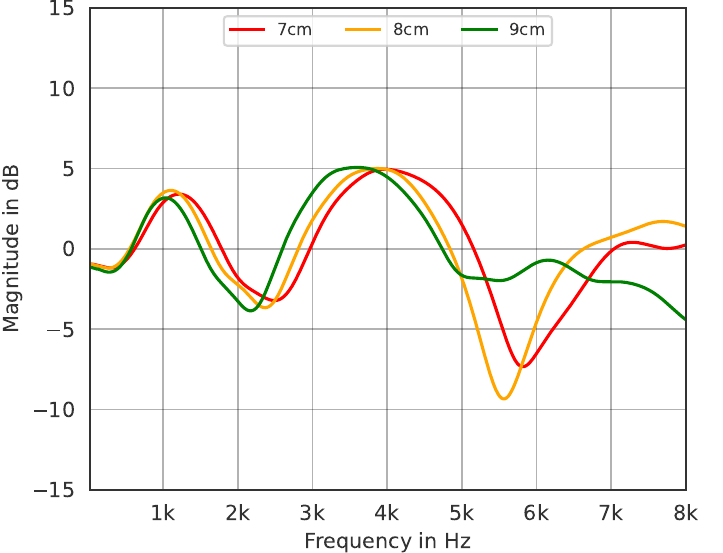}}
    \end{minipage}%

    \begin{minipage}{.33\textwidth}
        \centering
        \subfigure[Rigid sphere at varying distances.]{
            \includegraphics[trim=0.2cm 0.25cm 33cm 0.25cm, clip, scale=0.36]{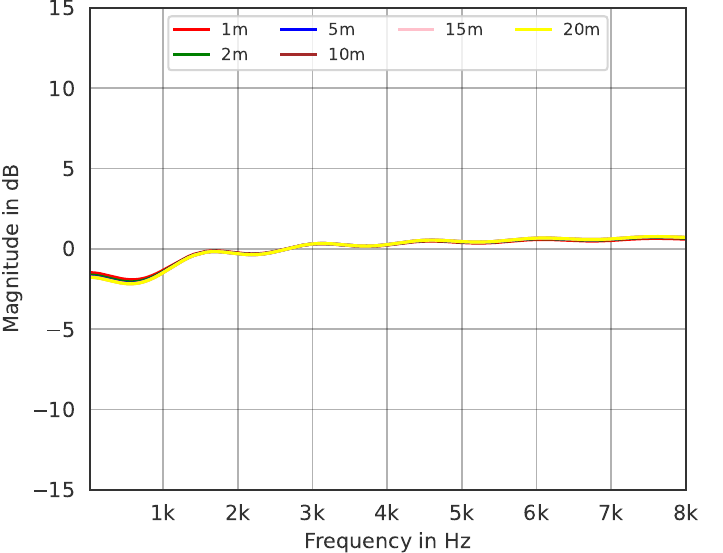}}
    \end{minipage}%
    \begin{minipage}{.33\textwidth}
        \centering
        \subfigure[Head at varying distances.]{
            \includegraphics[trim=0.2cm 0.25cm 33cm 0.25cm, clip, scale=0.36]{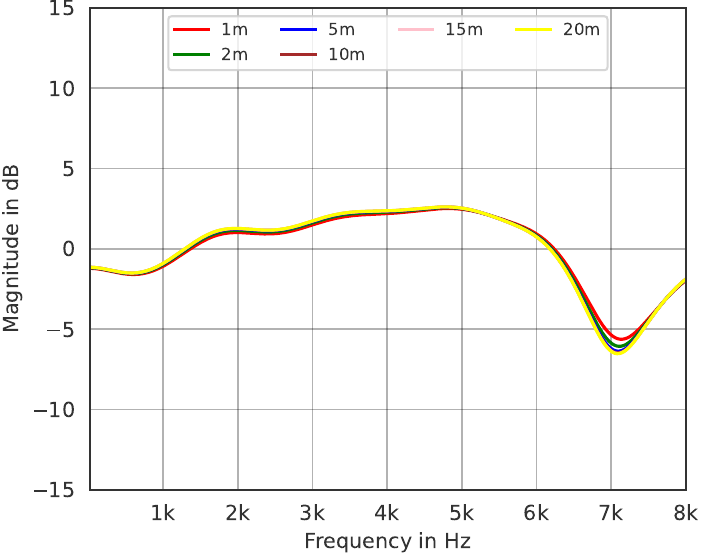}}
    \end{minipage}%
     \begin{minipage}{.33\textwidth}
        \centering
        \subfigure[Head-and-torso at varying distances.]{
            \includegraphics[trim=0.2cm 0.25cm 33cm 0.25cm, clip, scale=0.36]{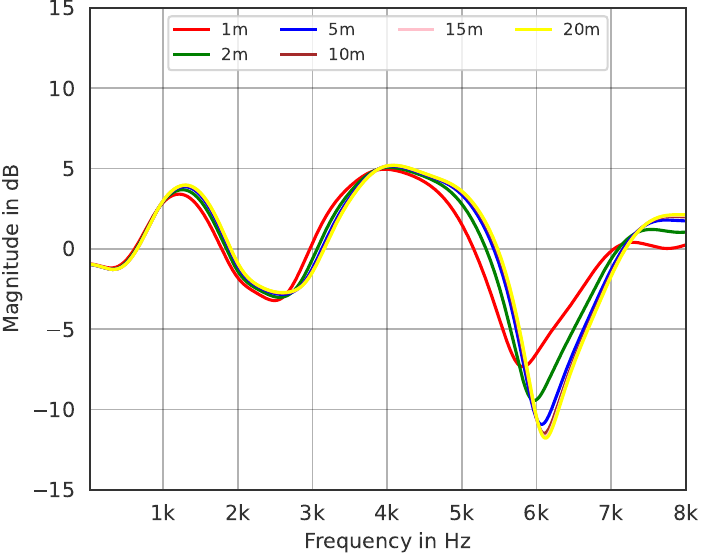}}
    \end{minipage}%
      
    \caption{ATFs for external speaker scenarios: (a)–(c) show a 3D model with a 7~cm head radius and a point source positioned 1~m away at various angles. (d)–(f) illustrate a 3D model with varying head radii and a point source positioned 1~m straight ahead. (g)–(i) depict a 3D model with a 7~cm head radius and a point source at varying distances.}
    \label{fig:fig10}
\end{figure*}

\begin{figure*}[hbt!]
    \centering
    \begin{minipage}{.33\textwidth}
        \centering
        \subfigure[Rigid sphere with varying caps.]{
        \includegraphics[trim=0.2cm 0.25cm 33cm 0.25cm, clip, scale=0.36]{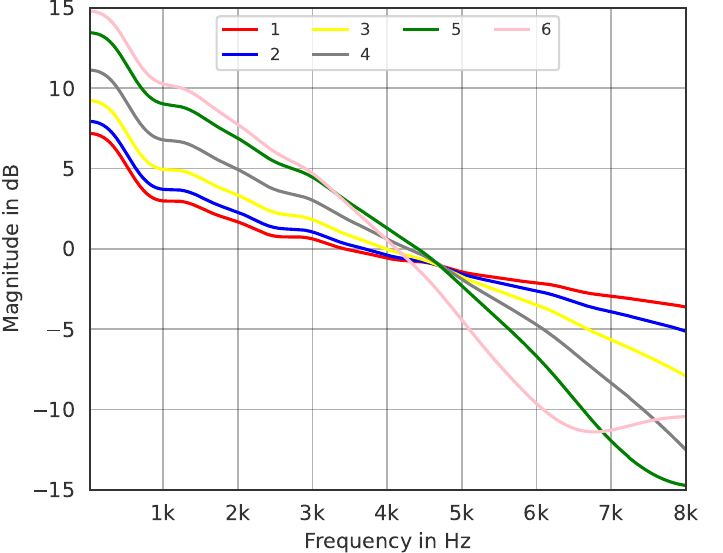}}
    \end{minipage}%
   \begin{minipage}{.33\textwidth}
        \centering
        \subfigure[Head with varying mouths.]{
        \includegraphics[trim=0.2cm 0.25cm 33cm 0.25cm, clip, scale=0.36]{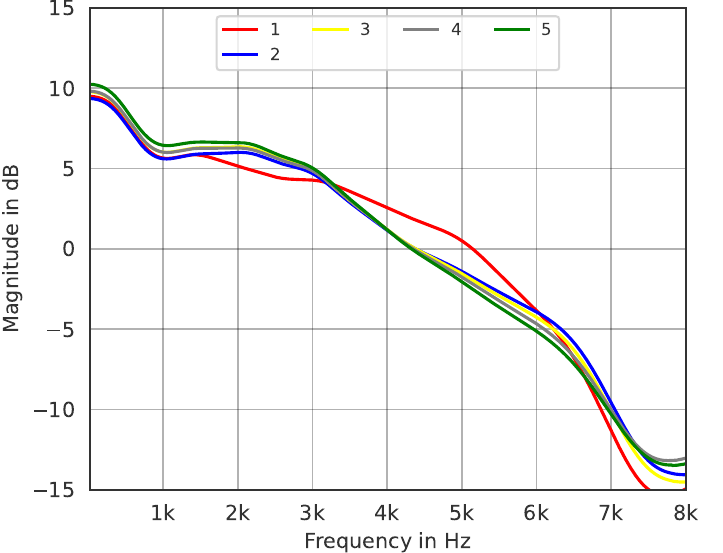}}
    \end{minipage}%
     \begin{minipage}{.33\textwidth}
        \centering
        \subfigure[{\fontsize{7.8}{7}\selectfont Head-and-torso with varying mouths.}]{
        \includegraphics[trim=0.2cm 0.25cm 33cm 0.25cm, clip, scale=0.36]{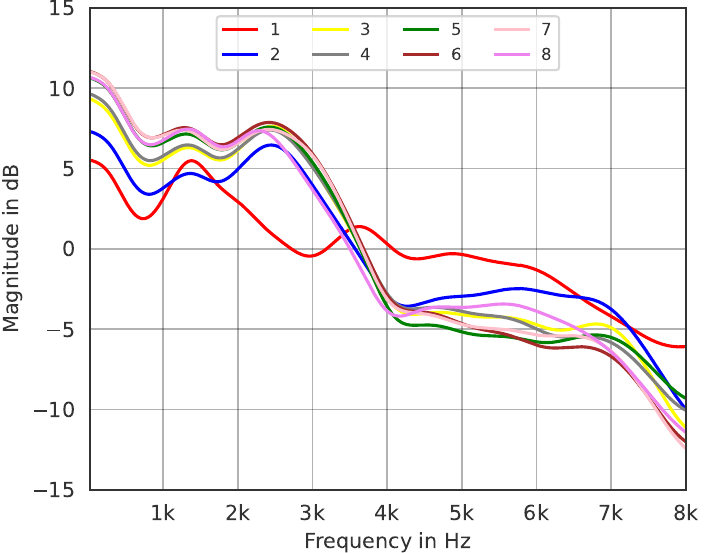}}
    \end{minipage}%

     \begin{minipage}{.33\textwidth}
        \centering
        \subfigure[Rigid sphere with varying radii.]{
        \includegraphics[trim=0.2cm 0.25cm 33cm 0.25cm, clip, scale=0.36]{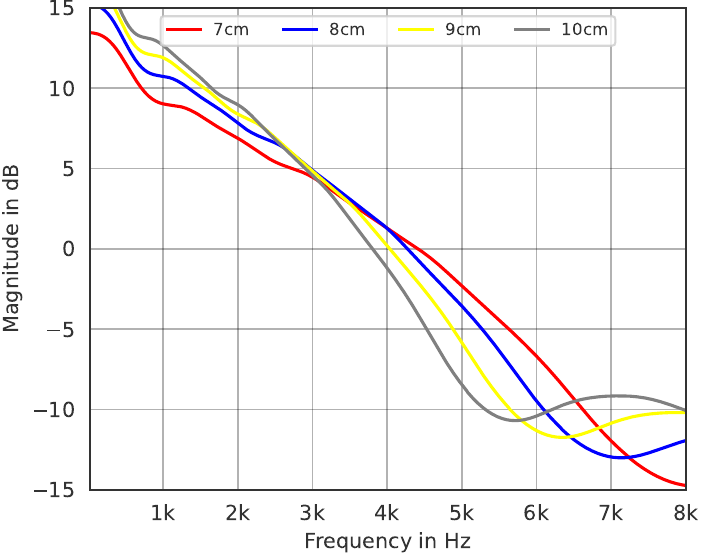}}
    \end{minipage}%
    \begin{minipage}{.33\textwidth}
        \centering
        \subfigure[Head with varying radii.]{
        \includegraphics[trim=0.2cm 0.25cm 33cm 0.25cm, clip, scale=0.36]{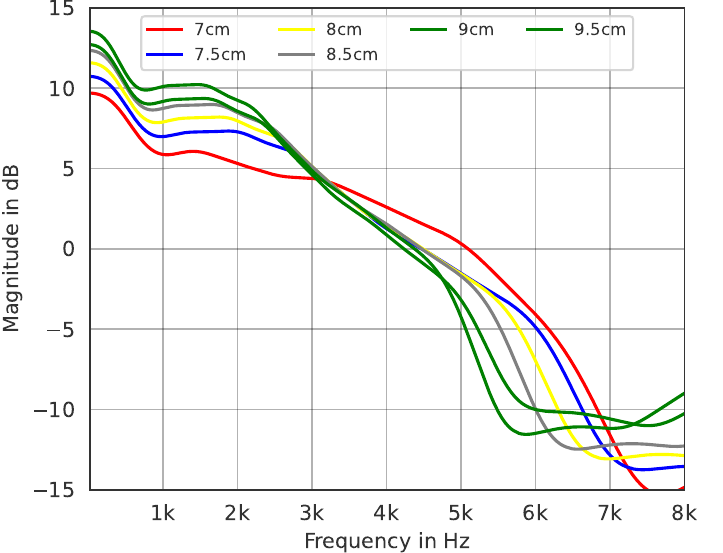}}
    \end{minipage}%
     \begin{minipage}{.33\textwidth}
        \centering
        \subfigure[Head-and-torso with varying radii.]{
        \includegraphics[trim=0.2cm 0.25cm 33cm 0.25cm, clip, scale=0.36]{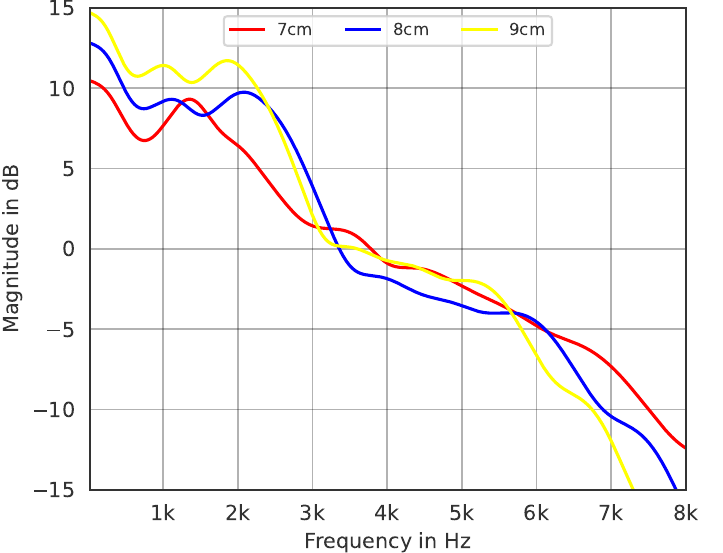}}
    \end{minipage}%
    
    \caption{ATFs for own voice scenarios: (a)–(c) show a 3D model with a 7~cm head radius, with varying mouth angles, where the number indicates increasing angles. (d)–(f) illustrate a 3D model with varying head radii. }
    \label{fig:fig23}
\end{figure*}

\subsection{Numerical Modeling Approach}
\label{sec:numerical_modeling}
We used \textit{Mesh2HRTF}~\cite{ziegelwanger2015mesh2hrtf, kreuzer2022mesh2hrtf}, an open-source software package for the numerical calculation of HRTFs to generate ATFs from 3D mesh models. We utilized its implementation of the multilevel fast multipole method (ML-FMM-BEM), a combination of the multilevel fast multipole method (ML-FMM)~\cite{chen2008formulation} and boundary element method (BEM)~\cite{gaul2013boundary}. The 3D model meshes of rigid sphere (see Fig.~\ref{fig:fig15}(a)) and human models (see Fig.~\ref{fig:fig15}(b), (c)) were created using Blender~\cite{blender} and post-processed using MeshInspector~\cite{meshinspector}.

\begin{table*}[!b]
	\caption{Description of the ATF dataset. (*) indicates the number of variations considered for each parameter.}
	\centering
    \renewcommand{\arraystretch}{1.2}
\begin{tabular}{ w{c}{1.6cm}w{c}{1.64cm}w{c}{1.74cm}w{c}{1.85cm}w{c}{1.46cm}w{c}{1.65cm}w{c}{1.2cm}w{c}{1.23cm}w{c}{1.6cm} }

    \toprule
	Model & Angles & Head radius & Distance & Mouth angle & Ear location & Total ATF & Own voice & Other speaker\\
	\midrule
	\multicolumn{9}{c}{Analytical approach} \\
	\midrule
	Rigid sphere   & 0$^\circ$-360$^\circ$ (360)   &7cm-10cm (50)& 0.5m-20m (100) &   5$^\circ$-20$^\circ$ (15) & 90$^\circ$-120$^\circ$ (20) & 153,516   &   76,757 &76,759\\
	\midrule
	\multicolumn{9}{c}{Numerical approach} \\
	\midrule
	Rigid sphere   & 0$^\circ$-360$^\circ$  (18)   &7cm-10cm (4)&   1m-20m (39)&   (6)& 90$^\circ$ &  2,544   &   48 &2,496\\
	Human head   &  0$^\circ$-360$^\circ$  (18)     &7cm-9.5cm (6)&   1m-20m (24)&   (5)&  90$^\circ$ & 2,414&   110 & 2,304\\
	Head \& torso   &  0$^\circ$-360$^\circ$  (18)     &7cm-9cm (3)&   0.5m-20m (25)&   (8)&  90$^\circ$ &   2,904&   336 & 2,568\\
	\bottomrule
\end{tabular}
\label{table:tab1}
\end{table*}
The rigid sphere mesh enables direct comparison with our analytical model and provides a foundational step for progressively fine-tuning the model toward more anatomically realistic head and torso representations. In Mesh2HRTF, radiation from vibrating elements were used to model the vibrating cap for the rigid sphere and the mouth for human models, while the external speaker was represented as a point source. The source code for inspecting SOFA files was modified, and customized evaluation grids were generated to suit our specific requirements. The simulation scenarios for other speakers and own voice were generated as illustrated in Figs.~\ref{fig:fig3}(a) and \ref{fig:fig3}(b), respectively.
Figure~\ref{fig:fig24} compares analytical and numerical simulation results for the rigid sphere model, showing close agreement between the two approaches.

\begin{figure}[!hbt]
    \centering
   \includegraphics[trim=0.2cm 0.25cm 33cm 0.25cm, clip, scale=0.4]{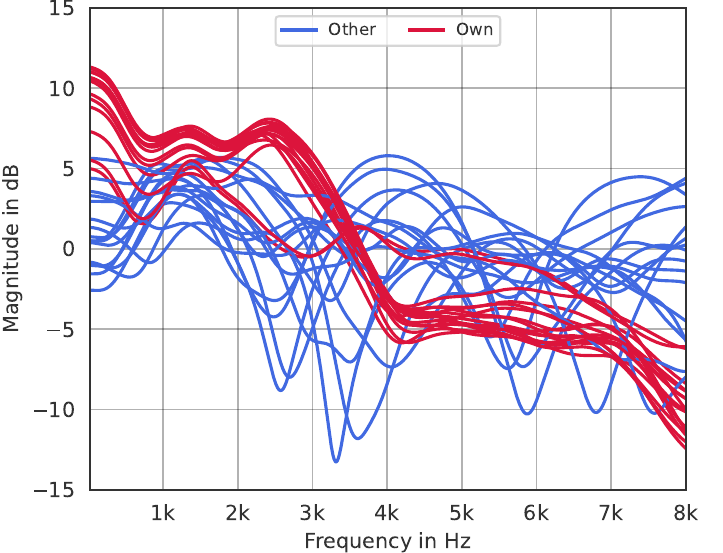}
    \caption{Comparison of own-voice and external-speaker ATFs at various angles using the head-and-torso model.}
    \label{fig:fig29}
\end{figure}

The ATFs were generated for three models of increasing anatomical complexity: a rigid sphere, a human head, and a head-and-torso configuration. These ATFs account for variations in speaker angles, distances, and anatomical dimensions. In addition, ATFs were generated for elevation angles of $+5^\circ$ and $-5^\circ$ in head-and-torso models to account for slight vertical speaker offsets. Figure~\ref{fig:fig10} summarizes the ATFs for external-speaker scenarios. Subplots (a)-(c) demonstrate how directional angle influences ATFs across models, with the rigid sphere showing smoother responses and the head-and-torso model exhibiting stronger spectral notches and directional dependence, especially above 3 kHz. Subplots (d)-(f) highlight the impact of changing head radius, where increases in head size lead to more pronounced high-frequency attenuation, particularly evident in the human head and head-and-torso cases. Subplots (g)–(i) show that increasing distance reduces fluctuation strength in the ATF, though structural model differences persist.

Figure~\ref{fig:fig23} focuses on the ATFs for the own voice condition. Unlike external sources, these responses exhibit a steep spectral slope, particularly in the head-and-torso model, indicating strong near-field characteristics and spatial filtering. Furthermore, Fig.~\ref{fig:fig29} directly compares own-voice and external-speaker ATFs using the head-and-torso model. This comparison further underscores that own-voice ATFs maintain steep spectral slopes with relatively consistent patterns, while external speaker ATFs show greater angular variability and flatter responses.

\vspace{-1mm}
\subsection{Own voice detection algorithm}
\label{sec:ovd_algorithm}
We apply transfer functions derived from both analytical and numerical simulation models to a clean speech wave in order to replicate how a hearing aid user perceives incoming sound. To incorporate these ATFs into the speech, we first extract the Short-Time Fourier Transform (STFT) from raw audio data, and the extracted STFT is then modified by applying the ATFs, simulating how sound propagates to the hearing aid user's ear.  Following this, a log-mel spectrogram is derived from the transformed STFT and used as input to the classifier. 

We formulate OVD as a segment-level binary classification task, where each input segment is assigned a single label (own-voice or external-speaker). We adopt a transformer-based model with temporal gate pooling~\cite{kawano2023effective}, which aggregates frame-level features across the segment into a single decision. This temporal aggregation improves robustness to phonetic variability and short-term spectral fluctuations while leveraging spatial–spectral cues embedded by the ATFs. We emphasize that the present work evaluates offline segment-level classification.

\section{Experimental setup}
In this section, we describe the dataset used for training and evaluation, the model architecture and hyperparameters, and the fine-tuning strategies employed.
\begin{table}[!t]
	\caption{Distribution of ATF dataset into training, validation, and testing sets for classes 0 (Other) and 1(Own).}
	\centering
    \renewcommand{\arraystretch}{1.2}
	\begin{tabular}{lp{0.5cm}p{0.5cm}p{0.4cm}p{0.4cm}p{0.4cm}p{0.4cm}}
		\toprule
		\multirow{2}{*}{ATF Dataset} & \multicolumn{2}{c}{Training} & \multicolumn{2}{c}{Validation} & \multicolumn{2}{c}{Testing} \\
		\cline{2-7}
		& 1 & 0 & 1 & 0 & 1 & 0 \\
		\midrule
		Rigid sphere (Analytical) & 69,180 & 69,181 & 3,452 & 3,452 & 4,125 & 4,126 \\
		Rigid sphere (Numerical) & 33 & 1,747 & 5 & 250 & 10 & 499 \\
		Head model (Numerical) & 77 & 1,612 & 11 & 231 & 22 & 461 \\
		Head \& torso (Numerical) & 235 & 1,797 & 33 & 257 & 68 & 514 \\
		\bottomrule
	\end{tabular}
	\label{table:tab2}
\end{table}

\subsection{Dataset}
VoxCeleb1 dataset~\cite{nagrani2017voxceleb} was used as the source of clean speech waveforms for generating ATF-augmented training data. The dataset contains 153,516 utterances, which are randomly assigned to the own voice or external speaker class (50\% each) prior to ATF-based transformations. The ATF dataset was generated using both analytically derived ATFs and numerically simulated ATFs. For the numerically simulated ATFs, each ATF class (external-speaker and own-voice) was split into training (70\%), validation (10\%), and testing (20\%), ensuring that test ATFs are not seen during training. The detailed ATF configuration parameters and dataset sizes are provided in Table~\ref{table:tab1}, and the dataset split is summarized in Table~\ref{table:tab2}. For each speech utterance, an ATF is randomly sampled from the corresponding split and applied to generate the ATF-augmented waveform.

To assess noise robustness, the MUSAN dataset~\cite{snyder2015musan}, which includes a diverse set of noises such as music, speech, and ambient sounds, was used to augment the training data by adding background noise at various SNR levels (0, 5, 10, 15, and 20 dB). To assess generalization to unseen speech content and speakers, we evaluate the model on LibriSpeech~\cite{panayotov2015librispeech}, which is not used during training. OVD test samples are generated by applying the same ATF augmentation procedure using held-out ATFs from the test split. The resulting LibriSpeech test set contains 4,125 utterances per class.

We also included real-world recordings from a hearing-aid prototype to assess the model’s robustness under practical acoustic conditions. Three participants provided informed consent prior to data collection. Recordings were obtained using two prototype units of the same hearing-aid device design connected to a Mackie ProFX16 audio mixer in a controlled acoustic room. For the own-voice condition, participants wore the device and read pre-defined sentences to ensure consistent speech content. For the external-speaker condition, participants continued wearing the device while a second person spoke in the same room, representing external speech in a realistic conversational scenario. The estimated SNR of the recordings was approximately 23~dB. For evaluation, 250 non-overlapping one-second segments were extracted, evenly distributed between the two classes. A randomly selected, disjoint subset of 50 segments was used for calibration to estimate statistics for domain-adaptation compensation, while the remaining segments were used exclusively for testing. Due to the limited size of the real-world dataset, this evaluation primarily serves as a proof-of-concept validation of simulation-to-real-world transfer.

In addition to real-world recordings, we evaluated the proposed approach on a publicly available measured-ATF hearing-aid dataset~\cite{lopez2019keyword}. In this dataset, clean speech signals from the Google Speech Commands dataset~\cite{warden2018speech} are convolved with ATFs measured from a hearing aid worn by a human subject, yielding both own-voice and external-speaker conditions. The measured transfer functions consist of 48 external-speaker HRTFs measured at different azimuths and one own-voice transfer function. Although the dataset provides multi-microphone recordings, only the front-microphone channel is used to match the single microphone setting considered in this work.

\vspace{-3mm}
\subsection{Model}
We followed the standard training approach from~\cite{kawano2023effective}, which employs a conformer-based encoder~\cite{gulati2020conformer}, and reduced the number of parameters (0.9M) to optimize computational efficiency while maintaining classification performance. The model is configured with a hidden size of 128, a feed-forward hidden size of 512, 4 attention heads, and a single encoder layer. In preprocessing, raw audio data of 16~kHz sample rate is segmented into fixed 15-second segments. The STFT is computed using a 25 ms window with a 10 ms stride, and an 80-channel log-mel spectrogram is used as input for the ML model.

\vspace{-3mm}
\subsection{Training methodology}

The model was trained using a progressive adaptation training strategy~\cite{bengio2009curriculum}, where the dataset complexity gradually increases to improve its performance in OVD. This approach follows a structured, multi-phase training process:

\begin{enumerate}
\item Training with analytical ATFs: Establishes a foundational understanding of sound propagation characteristics by allowing a wider range of parameter variations, including angles, distances, and head radii.
For training, an AdamW optimizer with
a learning rate of 1e-4 and a linear-warmup-and-cosine-decay scheduler were employed, conducting
training across a maximum of 50 epochs with a batch size of 64.

\item Fine-tuning with numerical ATFs: The model is fine-tuned using simulated ATFs, progressively transitioning from a rigid sphere model $\rightarrow$  human head model $\rightarrow$ head-and-torso model, allowing it to incrementally adapt to more complex practical applications. Each fine-tuning stage was run for 5 epochs. We employed two distinct fine-tuning strategies: (1) fine-tuning only the last classification layer and (2) fine-tuning both the feed-forward and classification layers~\cite{barcovschi2023comparative}.
\end{enumerate}

To assess robustness under noisy conditions, the model was trained using simulated speech data mixed with noise. Specifically, 70\% of the training data consisted of clean speech, while the remaining 30\% was randomly mixed with noise samples from the MUSAN dataset~\cite{snyder2015musan} at varying SNR levels of 0, 5, 10, 15, and 20~dB.

\vspace{-2mm}
\subsection{Test-time feature compensation}
When evaluating the model on real-world hearing-aid recordings, a distribution mismatch arises because the model is trained on simulated ATF-augmented speech. To mitigate this mismatch, we apply a lightweight test-time feature compensation. Let $X$ denote the log-Mel spectrogram extracted from a real-world speech segment. Device-dependent spectral coloration introduced by the hearing aid~\cite{kates2008digital} is first compensated using a white-noise reference recording, yielding $\tilde{X} = X - \mu_W$, where $\mu_W$ is the mean log-Mel spectrum of recorded white noise. To align the feature distribution with the simulated training domain, we apply a lightweight test-time statistic matching step inspired by correlation-alignment approaches in unsupervised domain adaptation (e.g., CORAL~\cite{sun2016deep}). Specifically,
\begin{equation}
X' = (\tilde{X} - \mu_R) \odot \frac{\sigma_S}{\sigma_R} + \mu_S ,
\label{eq:domain_adaptation}
\end{equation}

where $\mu_R$ and $\sigma_R$ are the per-frequency mean and standard deviation estimated from a disjoint calibration subset of the evaluation data, and $\mu_S$ and $\sigma_S$ are the corresponding statistics estimated from the simulated training data. The operator $\odot$ denotes element-wise multiplication. This compensation is applied only during inference and does not require any model fine-tuning.
\vspace{-2mm}
\subsection{Few-shot real-world data fine-tuning}
\label{sec:fewshot}
While the above compensation method enables simulation-to-real-world transfer, we also investigate few-shot model adaptation using a small set of real-world recordings. We use 50 labeled real-world segments for fine-tuning and keep the remaining segments as a held-out test set. The split is sentence-independent, with no sentence appearing in both the fine-tuning and test sets. To mitigate overfitting due to the limited data, additive background noise sampled from the MUSAN corpus is used for data augmentation, increasing the effective training set to 300 samples. During fine-tuning, the pretrained encoder parameters are frozen, while the feed-forward layers and classifier are updated for 5 epochs.

\section{Results}
\begin{table}[bt!]
\caption{
Test accuracy (\%) for models fine-tuned with different ATF configurations using two strategies: (i) classification layer only, and (ii) feed-forward layers with the classification layer. Evaluation is performed on the head-and-torso ATF dataset using utterances up to 15~s.
}
    \centering
    \renewcommand{\arraystretch}{1.2}
\begin{tabular}{lcm{2.4cm}m{2.4cm}}
        \toprule
        \multirow{2}{*}{Model fine-tuned on} & \multicolumn{2}{c}{Test Accuracy (\%) on Head \& Torso ATF dataset} \\
        \cline{2-3}
        & Last Layer Fine-Tuned & Feed-Forward + Last Layer Fine-Tuned \\
        \midrule
        + Sphere ATF & 83.78 $\pm$ 0.24 & 91.46 $\pm$ 0.20 \\
        + Head ATF & 90.54 $\pm$ 0.30 & 93.40 $\pm$ 0.19 \\
        + Head \& torso ATF & 91.04 $\pm$ 0.18 & 95.52 $\pm$ 0.20 \\
        \bottomrule
    \end{tabular}
    \label{table:test1}
\end{table}

\begin{table*}[!b]
\caption{
Test accuracy (\%) across different fine-tuning stages and ATF datasets, demonstrating generalization performance. The first section of the table reports results from the model trained on the analytical ATF dataset and tested on all ATF datasets. The second section presents results after progressive fine-tuning of both the feed-forward and classification layers. All models were evaluated using one-second speech segments.}
    \centering
    \renewcommand{\arraystretch}{1.2}
    \begin{tabular}{lcccc}
        \toprule
        \multirow{2}{*}{Model} & \multicolumn{4}{c}{Test Accuracy (\%)} \\
        \cline{2-5}
        & Analytical ATFs & Rigid Sphere (Mesh) & Head Model (Mesh) & Head \& Torso (Mesh) \\
        \midrule
        \multicolumn{5}{c}{Training on analytical ATFs} \\
        \midrule
        Baseline (Analytical ATFs) (50 epochs)  & 89.59 $\pm$ 0.14  & 83.69 $\pm$ 0.25  & 86.43 $\pm$ 0.15 & \textbf{85.12 $\pm$ 0.14}  \\
        \midrule
        \multicolumn{5}{c}{Feed-Forward + classification layer fine-tuning on numerical simulation ATFs } \\
        \midrule
        + Fine-tuned on Rigid Sphere (5 epochs) & -  & 84.42 $\pm$ 0.27  & 88.74 $\pm$ 0.08  & \textbf{86.87 $\pm$ 0.19}  \\
        + Fine-tuned on Head Model (5 epochs)   & -  & -  & 92.61 $\pm$ 0.17  &  \textbf{88.51 $\pm$ 0.17}\\
        + Fine-tuned on Head \& Torso (5 epochs)& -  & -  & -  & \textbf{90.02 $\pm$ 0.29}  \\
        \bottomrule
    \end{tabular}

    \label{table:test_combined}
\end{table*}

In this section, we present and analyze the experimental results. The OVD task is formulated as a segment-level binary classification problem, where each speech segment is assigned a single label (own-voice vs.\ external-speaker). VoxCeleb1 contains variable-length utterances; during training we use the full available utterance duration up to a maximum of 15~s to expose the model to diverse phonetic content. The conformer encoder supports variable-length inputs via padding or cropping with attention masking and produces frame-level embeddings that are aggregated using temporal gate pooling to obtain a single segment-level prediction. At test time, we additionally evaluate the same trained model on shorter one-second segments to assess robustness under lower-latency conditions and to approximate practical real-world usage scenarios.

\subsection{Training and progressive fine-tuning}
The model trained solely on analytically simulated ATFs already demonstrated strong generalization to simulation-based test sets, achieving 86\% accuracy on the head-and-torso dataset without any fine-tuning. Table~\ref{table:test1} presents the test accuracy for each stage of fine-tuning. Testing was performed on the head-and-torso ATF dataset, and the test set includes variable-length utterances up to 15 seconds. When fine-tuning only the classification layer with numerically simulated ATFs from the same sphere geometry, accuracy dropped slightly to 83.78\%, likely due to domain shift between analytical and numerical ATFs. This indicates that updating only the classifier is insufficient to bridge the gap between different modeling approaches. Progressively fine-tuning the classification layer alone with head-and-torso ATFs improved the accuracy to 91.04\%, surpassing the original baseline. However,  fine-tuning both the feed-forward and classification layers led to a substantial performance gain, achieving 95.52\% accuracy on the same dataset. These findings highlight the importance of internal feature representations in adapting to complex spatial conditions, and indicate that limiting fine-tuning to the classification layer constrains the model’s generalization capacity. Table \ref{table:test_combined} shows the test accuracy of models evaluated on one-second utterances. The final fine-tuned model achieved an accuracy of 90.02\% on the head-and-torso ATF dataset.

\begin{table}[!t]
\caption{
Robustness to different noise types and SNR levels. Test accuracy (\%) of the model trained on VoxCeleb1 with MUSAN noise augmentation, evaluated on both VoxCeleb1 and LibriSpeech using one-second utterances. The test ATF set is based on the Mesh2HRTF head-and-torso configuration.}
\centering
\renewcommand{\arraystretch}{1.2}
\begin{tabular}{lccc}
\toprule
Noise Type & SNR (dB) & VoxCeleb1 & LibriSpeech \\
\midrule
\multirow{6}{*}{Noise} 
& 0   & 75.86 $\pm$ 0.55  & 73.24 $\pm$ 0.49 \\
& 5   & 77.88 $\pm$ 0.34  & 75.55 $\pm$ 0.53   \\
& 10  & 79.45 $\pm$ 0.42  & 77.10 $\pm$ 0.57   \\
& 15  & 80.70 $\pm$ 0.35  & 78.28 $\pm$ 0.51   \\
& 20  & 82.30 $\pm$ 0.39  & 79.62 $\pm$ 0.49  \\
& 30  & 85.69 $\pm$ 0.16  & 82.79 $\pm$ 0.38  \\
\midrule
\multirow{6}{*}{Music} 
& 0   & 71.07 $\pm$ 0.36  & 68.53 $\pm$ 0.42 \\
& 5   & 74.26 $\pm$ 0.21  & 71.70 $\pm$ 0.47  \\
& 10  & 77.62 $\pm$ 0.18  & 74.91 $\pm$ 0.37  \\
& 15  & 81.00 $\pm$ 0.28  & 77.67 $\pm$ 0.30\\
& 20  & 83.69 $\pm$ 0.26  & 80.37 $\pm$ 0.45 \\
& 30  & 87.96 $\pm$ 0.15 & 84.83 $\pm$ 0.46  \\
\midrule
\multirow{6}{*}{Speech} 
& 0   & 81.23 $\pm$ 0.24  & 75.46 $\pm$ 0.39   \\
& 5   & 84.19 $\pm$ 0.30  & 78.73 $\pm$ 0.27  \\
& 10  & 86.59 $\pm$ 0.33  & 81.62 $\pm$ 0.27   \\
& 15  & 88.41 $\pm$ 0.23  & 83.89 $\pm$ 0.49   \\
& 20  & 89.65 $\pm$ 0.09  & 85.67 $\pm$ 0.38  \\
& 30  & 90.67 $\pm$ 0.13  & 87.93 $\pm$ 0.33  \\
\bottomrule
\end{tabular}
\label{table:noise_table}
\end{table}

\subsection{Noise augmentation}
To further assess the robustness of the proposed model, we trained it with background noise augmentation using the MUSAN dataset. Table~\ref{table:noise_table} reports the test accuracy across different noise types (Noise, Music, and Speech) and SNRs. The model shows reasonable resilience to noise, with performance decreasing gradually as SNR decreases. Furthermore, to assess generalization, we tested the model on the LibriSpeech dataset with added noise, a dataset not seen during training, and observed consistently high performance across all SNR levels. These results confirm the model’s ability to generalize well across datasets and remain robust in diverse acoustic conditions. All evaluations were conducted on the Mesh2HRTF head-and-torso ATF test set using one-second utterances.

\subsection{Baseline comparison and measured-ATF evaluation}

\begin{table*}[!b]
\caption{Comparison between the baseline ResNet and the proposed Conformer-small architecture on the measured-ATF dataset. Performance and complexity metrics are reported per 101-frame audio segment ($\approx$1~s), with execution latency measured on a host CPU (Intel Core i7, 2.90~GHz). MACs were estimated per input segment using framework profilers.}
\centering
\label{tab:ovd_with_training_data}
\renewcommand{\arraystretch}{1.2}
\setlength{\tabcolsep}{5pt} 
\begin{tabular}{l c c c c c c c} 
\toprule 
\multirow{2}{*}{Model} & \multicolumn{4}{c}{Performance (\%) on measured-ATF data} & \multicolumn{3}{c}{Computational Complexity} \\ 
\cmidrule(lr){2-5}\cmidrule(lr){6-8} 
& Own acc. & External acc. & Overall acc. & Macro-F1 & \#Params (K)/ Size (MiB) & Estimated MACs (M) & Execution Latency (ms) \\ 
\midrule 
ResNet~\cite{lopez2019keyword} & 97.52 & 80.81 & 93.15 & 90.75 & 238.5/ 0.91 & 1,002.49 & 8.40\\ 
Conformer-small & \textbf{98.77} & \textbf{89.94} & \textbf{96.46} & \textbf{95.41} & 307.2 / 1.17 & \textbf{17.61} & \textbf{2.98} \\ 
\bottomrule 
\end{tabular}
\end{table*}
To provide a direct baseline comparison and to isolate the contribution of the proposed simulation-based training strategy, we conduct three complementary evaluations. We use the ResNet-based single-microphone OVD model of López-Espejo et al.~\cite{lopez2019keyword} as the baseline because, to the best of our knowledge, it is the only published baseline directly comparable to our setting. This setting involves offline, segment-level, single-microphone own-voice/external-speaker detection using measured hearing-aid ATFs. In all comparisons in this section, Conformer-small denotes a reduced configuration of the proposed conformer model, selected to have a parameter count similar to the ResNet baseline~\cite{lopez2019keyword}. All models in these comparisons are trained and evaluated using one-second input segments to match the offline segment-level OVD setting.

The first evaluation compares the proposed Conformer-small model with the ResNet baseline on the measured-ATF benchmark. The second evaluation compares ResNet and Conformer-small under a controlled simulated-ATF setting, where both models are trained and evaluated using the same ATF-augmented Google Speech Commands data generated using our ATF pipeline. The third evaluation is a same-architecture comparison between Conformer-small trained only on measured ATFs and Conformer-small pretrained on simulated ATFs and then fine-tuned on measured ATFs. This final comparison isolates the benefit of the proposed simulated-ATF pretraining strategy from the effect of model architecture.

Table~\ref{tab:ovd_with_training_data} compares the performance and computational complexity of the proposed Conformer-small and the ResNet-based OVD baseline of López-Espejo et al.~\cite{lopez2019keyword} on the measured-ATF benchmark. The ResNet baseline is trained using speech data generated with 49 measured ATF conditions, whereas the proposed system is pretrained using speech data generated with 17{,}966 analytically simulated ATF conditions and then partially fine-tuned using speech data generated with the same 49 measured ATF conditions. The proposed system improves own-voice, external-speaker, and overall accuracy, with a particularly large gain in external-speaker accuracy, while requiring substantially fewer MACs and lower CPU execution latency. These results indicate a better accuracy-complexity trade-off for offline segment-level OVD, which is important for hearing-aid deployment. Recent lightweight and low-power neural speech-enhancement studies~\cite{yan2025lisennet,liu2026sse} further highlight the importance of computational efficiency for resource-constrained audio devices.

\begin{table}[!htb]
\centering
\caption{OVD performance on simulated-ATF Google Speech Commands dataset \cite{warden2018speech}. Both models use the same dataset split and are sequentially fine-tuned up to the head-and-torso ATF stage.}
\label{tab:ovd_sim_atf_comp}
\renewcommand{\arraystretch}{1.2}
\begin{tabular}{wl{1.5cm} wc{1cm} wc{1.1cm} wc{1.4cm} wc{1.3cm}}
\toprule
\multirow{2}{*}{Model} & \multirow{2}{*}{\#Params (K)} & \multicolumn{3}{c}{Accuracy (\%) on head \& torso ATF data} \\
\cline{3-5}
 & & Own & External & Overall \\
\midrule
ResNet \cite{lopez2019keyword}     & 238.5 & 98.24 & 83.18 & 94.30 \\
Conformer-small            & 307.2 & \textbf{98.38} & \textbf{91.80} & \textbf{96.66} \\
\bottomrule
\end{tabular}
\end{table}

As an additional controlled architecture comparison, Table~\ref{tab:ovd_sim_atf_comp} compares ResNet and Conformer-small when both models are trained and evaluated on the same simulated-ATF Google Speech Commands data generated using our ATF pipeline with 17{,}966 analytically simulated ATF variations. The results show that Conformer-small achieves comparable own-voice accuracy while providing substantially improved external-speaker and overall accuracy relative to the ResNet model. We note that the overall accuracy is influenced by class imbalance, with a larger proportion of own-voice samples, consistent with the evaluation protocol in~\cite{lopez2019keyword}.
\begin{table*}[!t]
\centering
\caption{Same-architecture comparison for evaluating the benefit of simulated-ATF pretraining. Both systems use Conformer-small, and all trainable parameters are updated during measured-ATF training/fine-tuning. FT denotes fine-tuning.}
\label{tab:same_architecture_ablation}
\renewcommand{\arraystretch}{1.2}
\setlength{\tabcolsep}{6pt}
\begin{tabular}{l l c c c c c c}
\hline
Split protocol & Training strategy &
Own acc. & External acc. & Overall acc. &
Macro-Precision & Macro-Recall & Macro-F1 \\
\hline
Speaker-disjoint &
Measured only &
99.10 & 95.69 & 98.21 & 97.95 & 97.40 & 97.67 \\

Speaker-disjoint  &
Sim. pretrain + measured FT &
\textbf{99.22} & \textbf{96.96} & \textbf{98.63} &
\textbf{98.35} & \textbf{98.09} & \textbf{98.22} \\

\hline
External-ATF-disjoint &
Measured only &
99.16 & 93.87 & 97.51 & 97.67 & 96.52 & 97.07 \\

External-ATF-disjoint &
Sim. pretrain + measured FT &
\textbf{99.40} & \textbf{96.84} & \textbf{98.60} &
\textbf{98.62} & \textbf{98.12} & \textbf{98.36} \\
\hline
\end{tabular}
\end{table*}

Finally, Table~\ref{tab:same_architecture_ablation} isolates the contribution of the proposed simulated-ATF pretraining strategy by holding the model architecture fixed. Both systems use the same Conformer-small architecture: one is trained only on measured ATFs, while the other is pretrained on simulated ATFs and subsequently fine-tuned on measured ATFs. Therefore, the performance difference reflects the benefit of simulated-ATF pretraining rather than a difference in model architecture. Simulated-ATF pretraining improves performance across both split protocols, with a more pronounced gain under the external-ATF-disjoint protocol, where external-speaker accuracy increases from 93.87\% to 96.84\%, indicating improved generalization to unseen ATFs.

\subsection{Real-world hearing-aid recording evaluation}
The model fine-tuned on simulated head-and-torso ATFs achieved 90.02\% accuracy on the corresponding simulated test set. When evaluated on the real-world recordings using compensation (Eq. \ref{eq:domain_adaptation}), the model attained 80\% accuracy, demonstrating effective simulation-to-real-world transfer despite the absence of fine-tuning. The evaluation was performed on non-overlapping one-second segments. For reference, an oracle class-conditional compensation achieves 86.5\% accuracy, representing an upper bound for the compensation-based approach.

\begin{figure}[!hbt] 
    \centering
   \includegraphics[ scale=0.33]{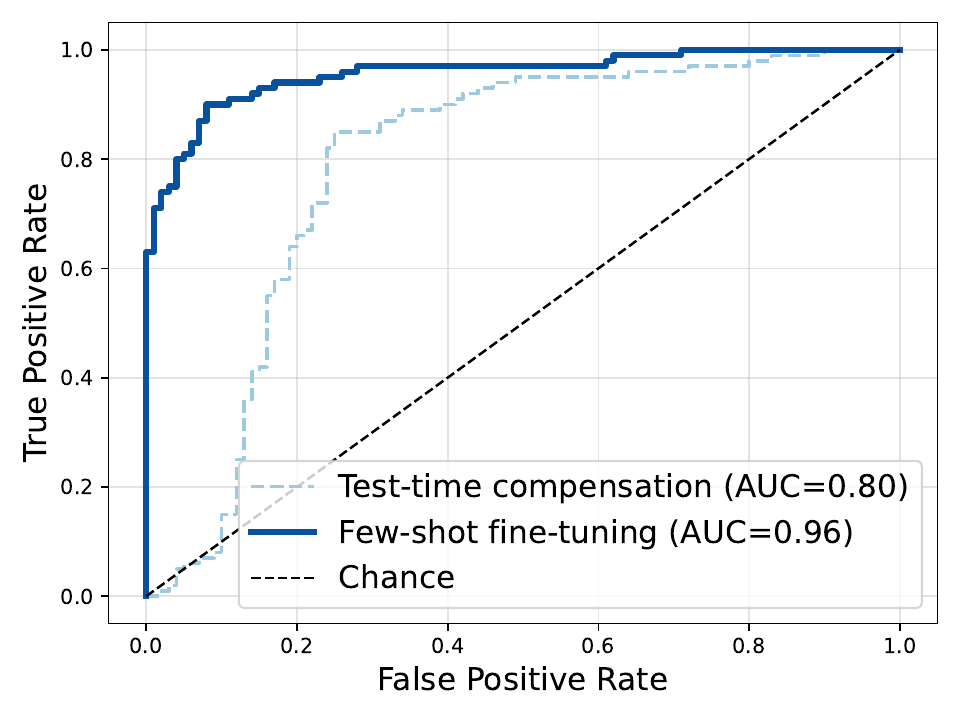}
   \caption{ROC curves for real-world test data.}
    \label{fig:fig31}
\end{figure}

Few-shot fine-tuning (see Section~\ref{sec:fewshot}) improves the real-world accuracy to 91\% without applying the compensation step during inference, indicating that limited model adaptation with a small set of real-world recordings can effectively reduce the simulation-to-real-world mismatch. For class-wise accuracy, the model obtains 92.0\% for the external-speaker class and 90.0\% for the own-voice class. 
The macro-precision, macro-recall, and macro-F1 scores are 91.06\%, 91.0\%, and 91.0\%, respectively. Figure~\ref{fig:fig31} shows ROC curves for the real-world hearing-aid recordings comparing test-time compensation and few-shot fine-tuning. Test-time compensation achieves an AUC of 0.80, while few-shot fine-tuning improves the detection performance to an AUC of 0.96. These results demonstrate the effectiveness of model adaptation and further support the viability of simulation-based training for OVD in hearing aids. The present work evaluates OVD in an offline, segment-level setting. A fully causal low-latency hearing-aid implementation is left for future work.
\subsection{Analysis of acoustic features for OVD}
While previous results demonstrate the model’s robustness, it remains important to understand the acoustic features the model relies on for OVD. In particular, loudness differences between the wearer’s voice and external speakers could be a simple cue that the model might exploit. To investigate this, we conducted experiments where the amplitude of the external speaker’s audio was artificially increased to simulate closer proximity. Despite these modifications, the model’s detection accuracy remained largely unchanged, achieving 89.90\% for the fine-tuned model on the head-and-torso dataset. These results indicate that the model does not rely solely on amplitude cues but instead leverages spatial acoustic features to discriminate between own voice and other speakers. This is crucial for real-world use, where speaker volume may vary significantly due to differences in speaking style or environment.

Additionally, repeating the one-second utterance to fill the 15-second input window did not improve accuracy compared to zero-padding, indicating that the model does not benefit from redundant temporal information. Finally, an ablation study where the ATF input was removed during testing resulted in a substantial drop in accuracy to 49.67\%, confirming that spatial cues are critical for the classification.

\section{Conclusion}
This study presents a simulation-based training framework for single microphone OVD in hearing aids. A conformer-based classifier was first trained on analytically derived ATFs to expose the model to a wide range of spatial configurations, and subsequently fine-tuned using numerically simulated ATFs with increasing anatomical realism. Unlike prior approaches that rely on multi-microphone arrays, speaker-dependent cues, or costly transfer-function measurements, the proposed method leverages simulated spatial propagation cues to distinguish own voice from external speakers. The model achieved 95.52\% accuracy on simulated head-and-torso test data using full-length utterances and maintained 90.02\% accuracy on one-second segments, demonstrating robustness to shorter input durations. Evaluation on the measured-ATF benchmark showed that simulated-ATF pretraining improves performance, particularly under unseen ATFs, indicating that exposure to more diverse simulated ATFs enhances the effectiveness and generalization of the model. Few-shot fine-tuning using a small set of real-world hearing-aid recordings achieved 91\% accuracy, indicating that the simulation-trained model can be effectively adapted using real-world data. Although the real-world validation is limited in scale, it provides an initial feasibility demonstration that models trained on simulated ATFs can generalize to the real world. Future work will focus on reducing model size, computational cost, and energy consumption for hearing-aid deployment, informed by recent advances in lightweight and low-power neural speech-processing models~\cite{yan2025lisennet,liu2026sse}, while also targeting shorter-window detection, causal real-time operation, and validation on larger and more diverse real-world hearing-aid datasets.

\section*{Acknowledgments}
Funding for this work was provided by GN Store Nord A/S.

\bibliographystyle{IEEEtran}
\bibliography{reference}
\begin{IEEEbiographynophoto}{Mathuranathan Mayuravaani} is a Ph.D. candidate at the School of Engineering and Computer Science, Victoria University of Wellington, New Zealand. She received her B.Sc. Hons (Computer Science) degree from the University of Jaffna, Sri Lanka.  Her research interests include speech signal processing and machine learning.
\end{IEEEbiographynophoto}
\begin{IEEEbiographynophoto}{W. Bastiaan Kleijn} is Professor at Victoria University of Wellington, New Zealand (since 2010) and a research scientist at Google (since 2011). He was a professor at TU Delft (2011-2021) and KTH Stockholm (1996-2014) and prior to that a Member of Technical Staff in the Research Division of AT\&T Bell Laboratories. He holds a PhD in Soil Science and an MSc in Physics from the University of California, Riverside, an MSEE from Stanford University, and a PhD in Electrical Engineering from TU Delft. He is Fellow of the IEEE (1999), the Royal Society of New Zealand (2021), and Engineering New Zealand (2024). 
\end{IEEEbiographynophoto}
\begin{IEEEbiographynophoto}{Andrew Lensen} is a Senior Lecturer and Programme Director of Artificial Intelligence at Victoria University of Wellington, New Zealand. He holds a B.Sc.(Hons) and a Ph.D. degree (2019) from Victoria University of Wellington. His current research interests include explainable AI,  interdisciplinary applications of AI, and the societal impacts of AI use in New Zealand. He is a Senior Member of the IEEE (2025).
\end{IEEEbiographynophoto}
\begin{IEEEbiographynophoto}{Charlotte Sørensen} is a Research Scientist at GN Hearing A/S, Denmark. She received the B.S. and M.Sc. degrees in Biomedical Engineering and Informatics, and the Ph.D. degree, from Aalborg University, Denmark. Her research interests include speech signal processing with emphasis on speech intelligibility prediction, measurement of speech intelligibility, and speech enhancement for hearing aid applications.
\end{IEEEbiographynophoto}


\end{document}